# Reconstructive comb spectroscopy: A single-pixel detection paradigm beyond dual-comb limitations

Dongxu Zhu[1]†, Zhuoren Wan[1,2]†, Xiaoshuai Ma[1], Chongyuan Shui[3], Yifan Yang[3], Yuan Chen[1], Mei Yang[1], Zijian Wang[1], Xiuxiu Zhang[1], Min Li[3], Hua Li[5], Kun Huang[1], Ming Yan[1,2]*, Yan Liang[3]*, Weiwei Cai[4]*, and Heping Zeng[1,2]

[1]State Key Laboratory of Precision Spectroscopy, and Hainan Institute, East China Normal University, Shanghai, China

[2]Chongqing Key Laboratory of Precision Optics, Chongqing Institute of East China Normal University, Chongqing 401120, China

[3]School of Optical-Electrical and Computer Engineering, University of Shanghai for Science and Technology, Shanghai 200093, China

[4]Key Lab of Education Ministry for Power Machinery and Engineering, School of Mechanical Engineering, Shanghai Jiao Tong University, Shanghai, 200240, China

[5]Key Laboratory of Terahertz Solid State Technology, Shanghai Institute of Microsystem and Information Technology, Chinese Academy of Sciences, Shanghai 200050, China

†These authors contributed equally.

* Corresponding author: *myan@lps.ecnu.edu.cn, yanliang@usst.edu.cn, cweiwei@sjtu.edu.cn*

**Abstract:**

Frequency comb spectroscopy has revolutionized broadband molecular fingerprinting with mode-defined resolution. While dual-comb spectroscopy stands as a dominant paradigm for high-resolution measurements, it relies on mutually coherent dual combs, and its applicability to non-cooperative sensing is limited by the requirement for phase-sensitive detection and controlled optical returns. Here, we introduce reconstructive comb spectroscopy, a fundamentally different paradigm that eliminates these constraints. By integrating a mode-programmable optical comb with a computational sensing scheme based on single-pixel detection, our method achieves picometer-level spectral resolution over a 10-nm (1.27-THz) instantaneous bandwidth, with single-photon sensitivity down to $10^{-4}$ photons per pulse, and compressed spectral acquisition at 2.5% sampling while maintaining reconstruction errors below 10%. We demonstrate robust performance through scattering media and from non-cooperative targets. These capabilities establish reconstructive comb spectroscopy as a new platform for gas sensing, with broad applicability in remote atmospheric monitoring, industrial leak detection, and standoff chemical-threat identification.

# Introduction

Spectroscopic gas sensing in hazardous or inaccessible environments demands non-contact measurements with broad spectral coverage and high resolution [1-4], enabling applications ranging from remote atmospheric monitoring [3] to standoff detection of industrial leaks and chemical threats [4]. Conventional dispersive and Fourier-transform spectrometers (Figs. 1a,b) are constrained by grating dimensions and optical path lengths, limiting spectral resolution and accuracy [5]. Frequency-swept laser spectroscopy [6] offers high precision but only narrow instantaneous bandwidths that impede simultaneous multi-gas detection. In contrast, broadband frequency comb spectroscopy has emerged as a powerful platform, offering unprecedented resolution and accuracy through mode-defined frequency lines [7, 8].

Among frequency-comb-based spectroscopic techniques, dual-comb spectroscopy (DCS) has established itself as a leading paradigm [9]. By harnessing two mutually coherent optical frequency combs with slightly offset repetition rates, DCS performs Fourier-transform spectroscopy through multi-heterodyne detection on a single-pixel detector (Fig. 1c), achieving rapid, high-resolution spectral acquisition without moving parts [10-13]. This capability has enabled applications in open-path greenhouse-gas monitoring [14-16] and multispecies trace-gas detection [17, 18]. However, extending DCS to remote and non-cooperative scenarios remains challenging. Reliable operation requires preserving mutual coherence at the detector, often necessitating cooperative targets such as calibrated reflectors [19, 20]. Furthermore, maintaining two phase-locked combs increases system complexity and cost, limiting deployment in compact and field-operable systems.

Alternatively, phase-insensitive comb spectroscopy using dispersive elements, such as a virtually imaged phased array (VIPA) [21] or multi-mode fibers [22], provides a camera-based approach, but struggles under photon-starved conditions [23], such as remote sensing of low-reflectivity targets or operation through highly scattering media (e.g., fog, dust, clouds). This gap highlights the need for a new spectroscopic paradigm that combines the metrological strengths of frequency combs with robustness to non-cooperative scenarios, while avoiding the cost and complexity of dual-comb systems.

Here, we introduce reconstructive comb spectroscopy as a complementary paradigm to existing frequency-comb spectroscopy. By leveraging a single mode-programmable comb and intensity-based computational sensing, our method encodes individual comb modes via a digital micromirror device (DMD) and reconstructs spectra from single-pixel measurements. Although DCS is computational in nature, relying on the algorithmic reconstruction of interferograms, it depends on coherent multi-heterodyne detection. In contrast, our reconstructive approach operates without phase-sensitive detection, thereby eliminating the need for dual-comb synchronization and cooperative targets. This architecture retains the core advantages of frequency comb spectroscopy, including absolute frequency traceability and high resolution, and additionally facilitates compressed sensing, single-photon sensitivity, and operation through scattering media. We demonstrate that reconstructive comb spectroscopy achieves performance comparable to established techniques while introducing a fundamentally different measurement paradigm that combines the metrological strengths of frequency combs with robustness to non-cooperative scenarios, positioning it as a versatile platform for next-generation remote gas sensing.

# Results

## Basic principle

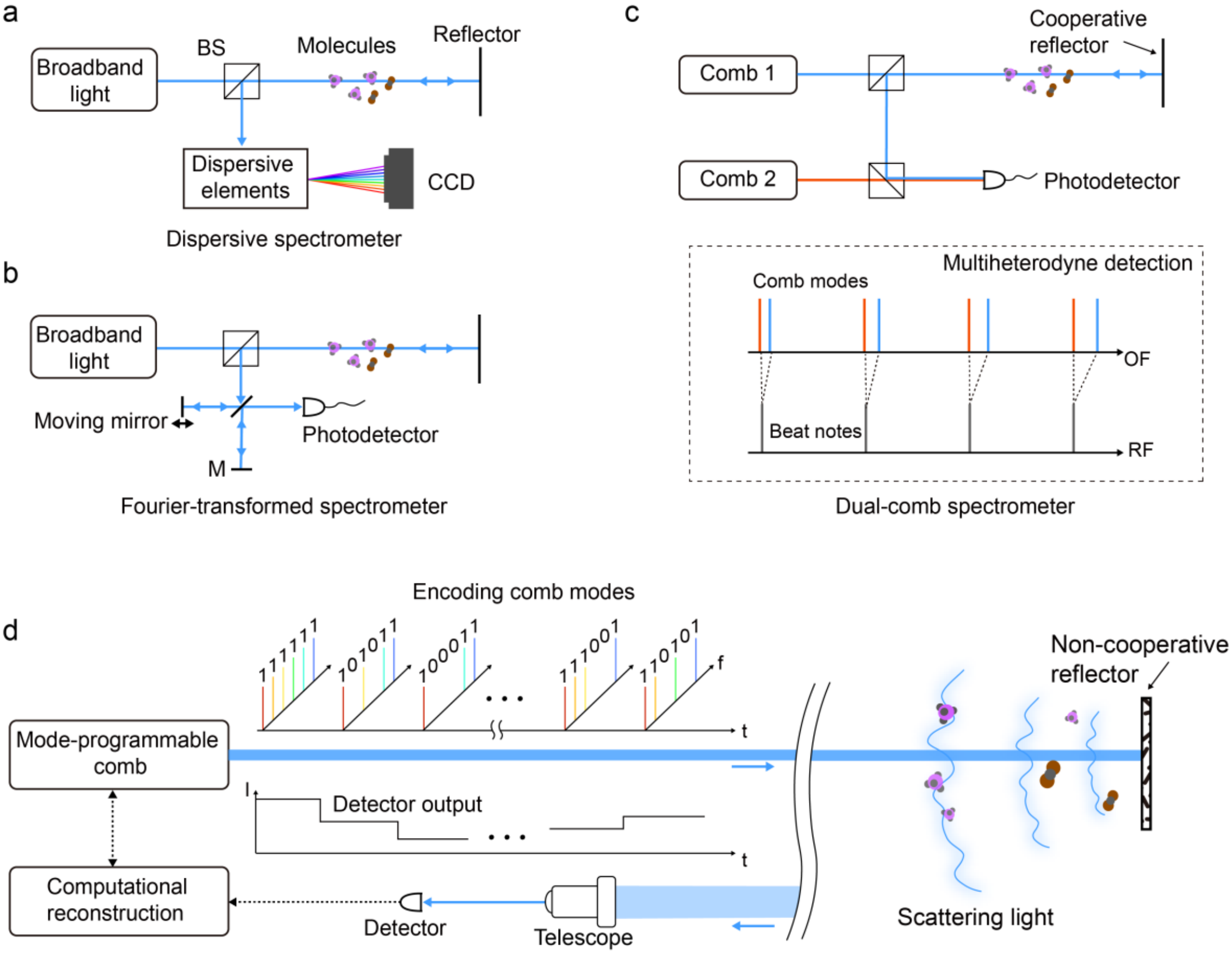

**Figure 1. Frameworks of frequency comb spectroscopy.**

**a,** Dispersive spectroscopic configuration.

**b,** Fourier-transform spectrometer based on a Michelson interferometer. M, fixed mirror.

**c,** Dual-comb spectroscopy based on multiheterodyne detection between two optical combs with slightly different mode spacings. High coherence between the two combs at the detector is required to resolve the dual-comb beat notes. OF, optical frequency; RF, radio frequency.

**d,** Mode-programmable frequency comb spectroscopy. In this scheme, by encoding the comb modes that probe the molecules and measuring only the intensity of the back-scattered light, a broadband, mode-resolved spectrum can be reconstructed computationally. I, intensity; t, time.

As illustrated in Fig. 1d, our approach employs a mode-programmable frequency comb that allows for arbitrary spectral encoding on a mode-by-mode basis, as each comb mode can be independently selected for computational spectroscopic sensing. By using a series of designed mode combinations, a broadband and high-resolution spectrum can be reconstructed from the intensity response of a single-pixel detector that measures back-scattered light. The underlying principle is analogous to that of computational imaging [24, 25]: the measured signal is an encoded projection of the unknown spectrum. Mathematically, the detection process can be expressed as

$$\mathbf{y}=\mathbf{H}\mathbf{s},$$

where **y** is the measured one-dimensional detection signal vector with $m$ elements and **H** is an $m\times n$ encoding matrix (with binary elements 0 and 1) defining the active comb-mode combinations, and $\boldsymbol{s}$ represents the normalized spectral coefficients at the comb mode frequencies $f_n$ (where $n$ is the comb mode index). The vector **s**, corresponding to the molecular spectrum under test, is computationally retrieved by solving the regularized inverse problem as

$$\hat{\mathbf{s}} = \underset{\mathbf{s}}{\operatorname{argmin}} \parallel \mathbf{Hs} - \mathbf{y} \parallel_2^2 + \boldsymbol{\gamma}\mathbf{R}(\mathbf{s})$$

where $\boldsymbol{R}(\boldsymbol{s})$ is a regularization term weighted by a coefficient $\boldsymbol{\gamma}$. Since efficient computational reconstruction algorithms (such as least-squares inversion [26] and compressive-sensing reconstruction [27]) are well established, we simply adopt these standard methods in our implementation. Further details are provided in Methods and Supplementary Note 1.

Notably, integrating computational sensing metrology with a mode-programmable comb offers several key advantages: (1) it retains the strengths of optical comb spectroscopy, including simultaneous broadband detection, high spectral resolution limited by the comb linewidth, and absolute frequency accuracy traceable to an atomic clock [9]; (2) it enables compressed spectroscopic sensing with substantially fewer measurements than the comb mode number; and (3) it operates without coherent detection or preset reflectors, making it particularly suitable for non-cooperative sensing scenarios.

## Experimental schematics

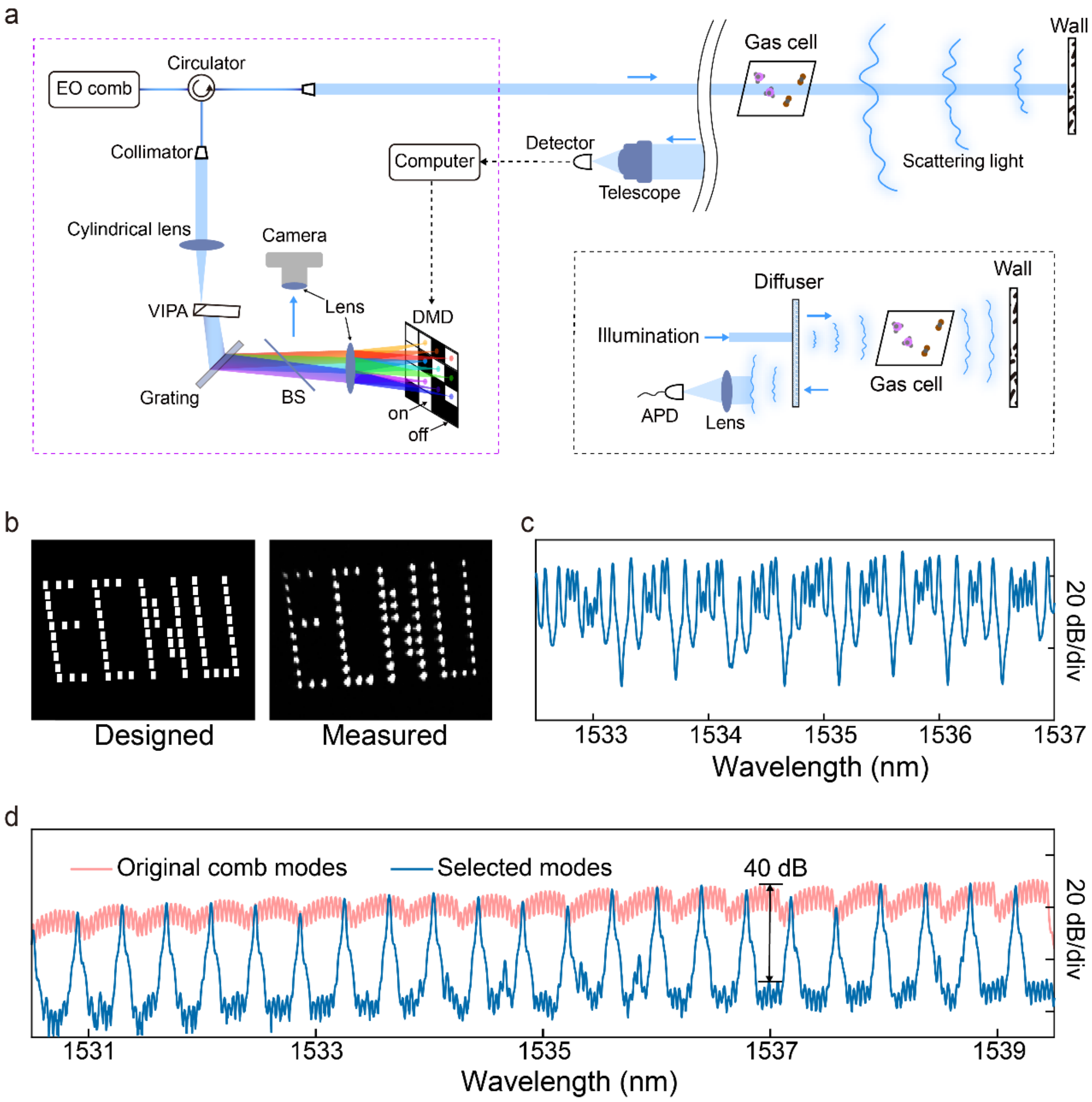

**Fig. 2 | Experimental details.**

**a,** Experimental setup. EO, electro-optic; VIPA, virtually imaged phased array; DMD, digital micromirror device; APD, avalanche photodiode.

**b,** Designed and measured results for a representative DMD coding pattern of the characters "ECNU" (abbreviation for East China Normal University).

**c,** Corresponding optical spectrum measured by a commercial spectral analyzer with 0.02 nm resolution.

**d,** Comb spectra before and after DMD encoding. The original comb mode spacing is 5 GHz (0.04 nm), and the encoding extinction ratio reaches a maximum of 40 dB.

Figure 2a depicts the experimental setup. The mode-programmable comb (outlined by the pink dashed box) comprises an electro-optic (EO) comb, a two-dimensional (2D) disperser, and a digital micromirror device (DMD) for comb-mode control. Further details are provided in Methods and Supplementary Fig. 1. In brief, the EO comb is seeded by a continuous-wave laser at 1550 nm with a specified linewidth below 0.1 kHz. The laser center frequency ($f_{cw}$) can be finely tuned within ±10 GHz via a piezoelectric transducer. The seed passes through an EO modulator, generating a comb with a mode spacing ($f_r$) of 5 GHz and a spectral coverage from 1520 to 1580 nm (189.74-197.23 THz; Supplementary Fig. 2). The stability of $f_r$ is directly inherited from a signal generator referenced to a rubidium clock.

To spatially control the comb modes, a virtually imaged phased array (VIPA; free spectral range: 60 GHz) and a transmission grating (966 lines $mm^{-1}$) jointly disperse the comb into a 2D frequency-space map, in which individual modes are projected to distinct spatial positions to form a frequency-resolved lattice. A lens (f=250 mm) focuses the dispersed comb onto the DMD (10 kHz switching rate), which digitally encodes the selected frequency components. To characterize the encoder, we capture the coded comb modes using a camera. As shown in Fig. 2b, the recorded image faithfully reproduces the designed letter patterns projected onto the DMD. The corresponding encoded spectrum, spanning approximately 1530 to 1540 nm as measured with an optical spectrum analyzer, is presented in Fig. 2c. This portion of the comb spectrum is selected for two reasons: (1) it exhibits relatively small comb-mode intensity variation (within ~10 dB), and (2) it best matches the bandwidths of the optical components in our system.

Figure 2d further demonstrates the simultaneous encoding of 228 comb modes: modes encoded as '1' preserve their original intensity, whereas those encoded as '0' are suppressed to the noise floor, achieving a maximum modulation contrast of 40 dB. These results verify the ability to manipulate the comb in a mode-by-mode manner, enabling programmable spectral encoding that underpins subsequent computational sensing and non-cooperative measurements.

For spectroscopic gas sensing, the encoded comb is launched into free space through a fiber-coupled collimator. The output power is below 30 mW, but can be further amplified using an erbium-doped fiber amplifier (EDFA) positioned either before or after the encoding stage. The collimated comb passes through a gas cell and then impinges on a concrete wall (reflectivity above 50%), which acts as an incoherent, diffusely scattering surface. The scattered light is collected by a telescope and focused onto a free-space detector, either an avalanche photodiode (APD) or a single-photon detector (SPD). The detector output is digitized (or photon-counted) and subsequently processed using the known encoding patterns to reconstruct the absorption spectrum of the target gas molecules.

### Spectroscopic validation

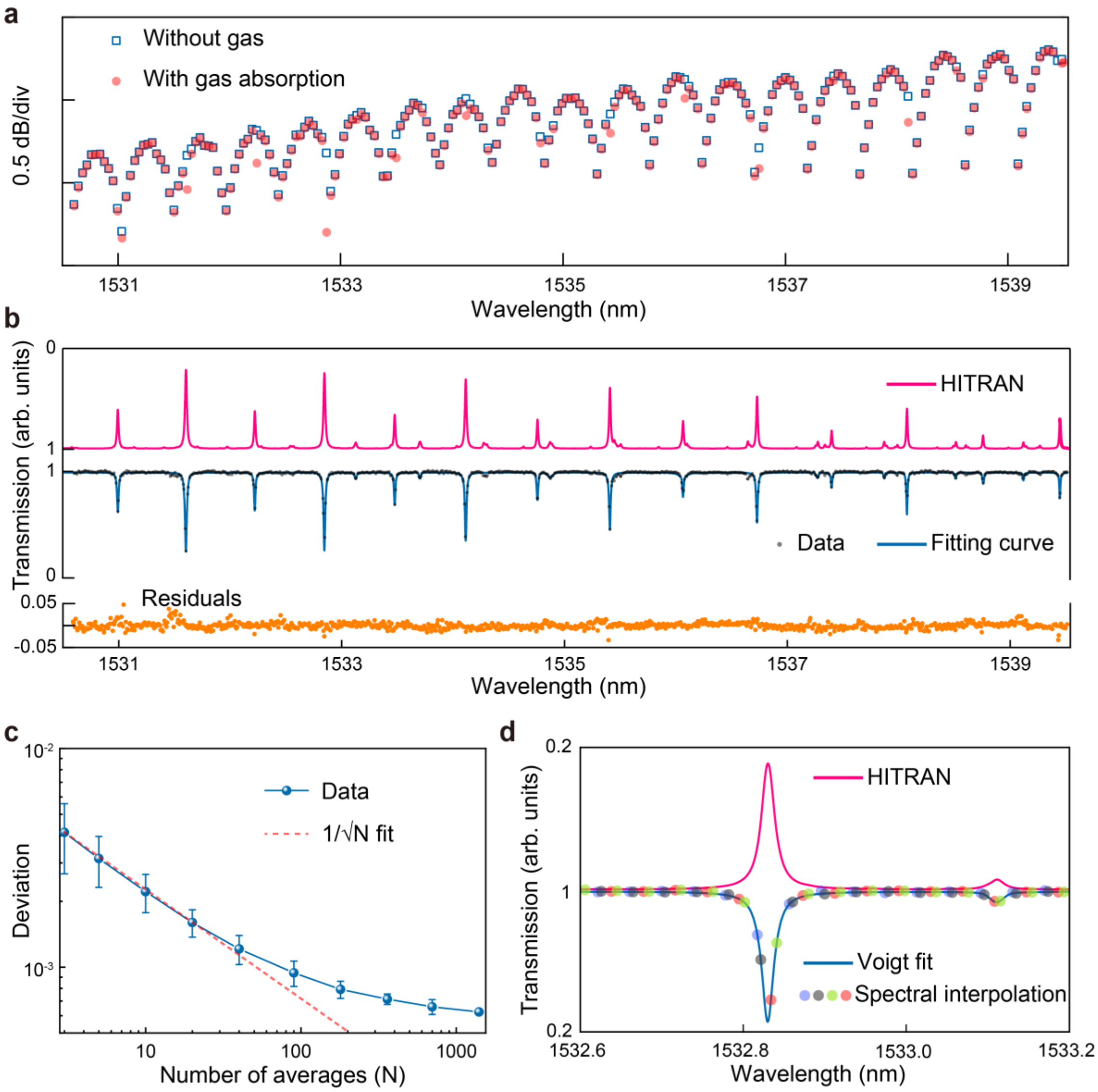

**Fig. 3 | Spectral results.**

**a,** Reconstructed spectra with and without gas absorption at comb-resolved resolution.

**b,** Normalized transmission spectrum (gray dots) fitted with Voigt profiles (blue), compared with a HITRAN simulation (pink). The residuals (orange points) exhibit a standard deviation below 1%.

**c,** Standard deviation as a function of the number of averages.

**d,** Interpolated absorption spectrum together with a simulation curve. The sampling points are spaced by 1.25 GHz (9.6 pm). For these measurements, a 13.5-cm-long gas cell was filled with 10% $^{12}C_2H_2$ and 90% $N_2$ at $3.5\times10^4$ Pa and 275 K.

We organize the results into four progressively challenging demonstrations: (i) spectroscopic validation to verify measurement accuracy, (ii) compressed sensing to demonstrate data efficiency, (iii) spectroscopic sensing through scattering media to evaluate robustness, and (iv) single-photon sensitivity to establish the photon-starved detection limit. We first validate the method by measuring the absorption spectrum of acetylene ($C_2H_2$). Figure 3a shows the reconstructed spectra obtained with and without the gas present. Each data point corresponds to the intensity or power of a single comb mode, with adjacent modes separated by $f_r$=5 GHz. The frequency axis is calibrated using the measured comb center frequency $f_{cw}$ (wavemeter accuracy ±10 MHz) together with the $f_r$. The reconstruction uses a differential Hadamard scheme that requires $2\times n$ encoding patterns for $n$ comb modes (see Methods). With a DMD switching rate of 10 kHz and $n$ = 256, the total acquisition time for the measurements is 51.2 ms. Note that points exceeding the total mode count (i.e., 228) are zero-padded (see Methods).

Figure 3b shows the transmission spectrum (black dots) after background normalization (see Methods), together with a Voigt fit (blue). For comparison, we also plot a simulated spectrum (pink) generated using parameters from the HITRAN 2020 database. The measured data agree well with the simulation, with the residuals (orange) exhibiting a standard deviation (SD) of 0.7%. The SD is further reduced by spectral averaging, following the expected $\sqrt{N}$ scaling for averaging numbers $N < 50$, as shown in Fig. 3c. Beyond this point, the improvement saturates due to mismatches between the simulated model and practical experimental conditions.

Meanwhile, spectral interpolation further enhances the resolution beyond the intrinsic comb mode spacing. Figure 3d shows interpolated data for the $\nu_1+\nu_3$ band P(13) transition of $^{12}C_2H_2$, achieving a resolution of 1.25 GHz (or 9.6 pm) through stepwise tuning of $f_{cw}$. This resolution is adequate for gas sensing under atmospheric pressure [6]. Nevertheless, data acquired with finer tuning steps, and thus higher resolution (100 MHz or 0.8 pm), are presented in Supplementary Fig. 3, with the corresponding theoretical resolution limits discussed in Supplementary Note 2.

## Compressed computational sensing

**Fig. 4 | Spectral results for Compressed sensing.**

**a,** Comb spectra reconstructed under different compression ratios using the TVAL3 algorithm. The point spacing is 1.25 GHz or 9.6 pm.

**b,** Zoom-in view of the dashed region in **a**.

**c,** Standard deviation (SD) between the uncompressed and compressed data as a function of the compression ratio.

Following validation of the foundational spectral reconstruction capability, we leverage a defining advantage of computational spectroscopy, which is its ability to recover *n* spectral elements from fewer than *n* measurements, enabling data compression and reduced acquisition time. We demonstrate this capability by reconstructing compressed comb-mode spectra using TVAL3 (a MATLAB-based total-variation minimization algorithm that preserves sparse spectral features via an ADMM optimization framework [28]). Meanwhile, the comb modes are encoded using Walsh codes to generate the compressed measurements.

Figure 4a presents the reconstructed comb spectra (4×228 modes with a 1.25 GHz mode spacing) at various compression ratios, with each point representing a comb-mode peak. As shown in the zoom-in view (Fig. 4b), the mode center frequencies remain well preserved across all ratios. The uncompressed (100%) spectrum is reconstructed from 2×1024 (= 2048) measurements acquired over 204.8 ms. The deviations in comb-mode intensities between the uncompressed and compressed spectra are also plotted in Fig. 4b. We quantify the SDs of these intensity discrepancies as a function of compression ratio (Fig. 4c). Within a 10% mode-intensity SD, we achieve a 2.5% compression ratio, corresponding to a 40-fold reduction in acquisition time (i.e., ~5 ms). Compressed spectra including gas absorption features are provided in Supplementary Fig. 4. Moreover, compressed spectra obtained using advanced adaptive search algorithms, which are particularly effective for sparse spectra with few peaks, are shown in Supplementary Fig. 5.

**Spectroscopic sensing through scattering media**

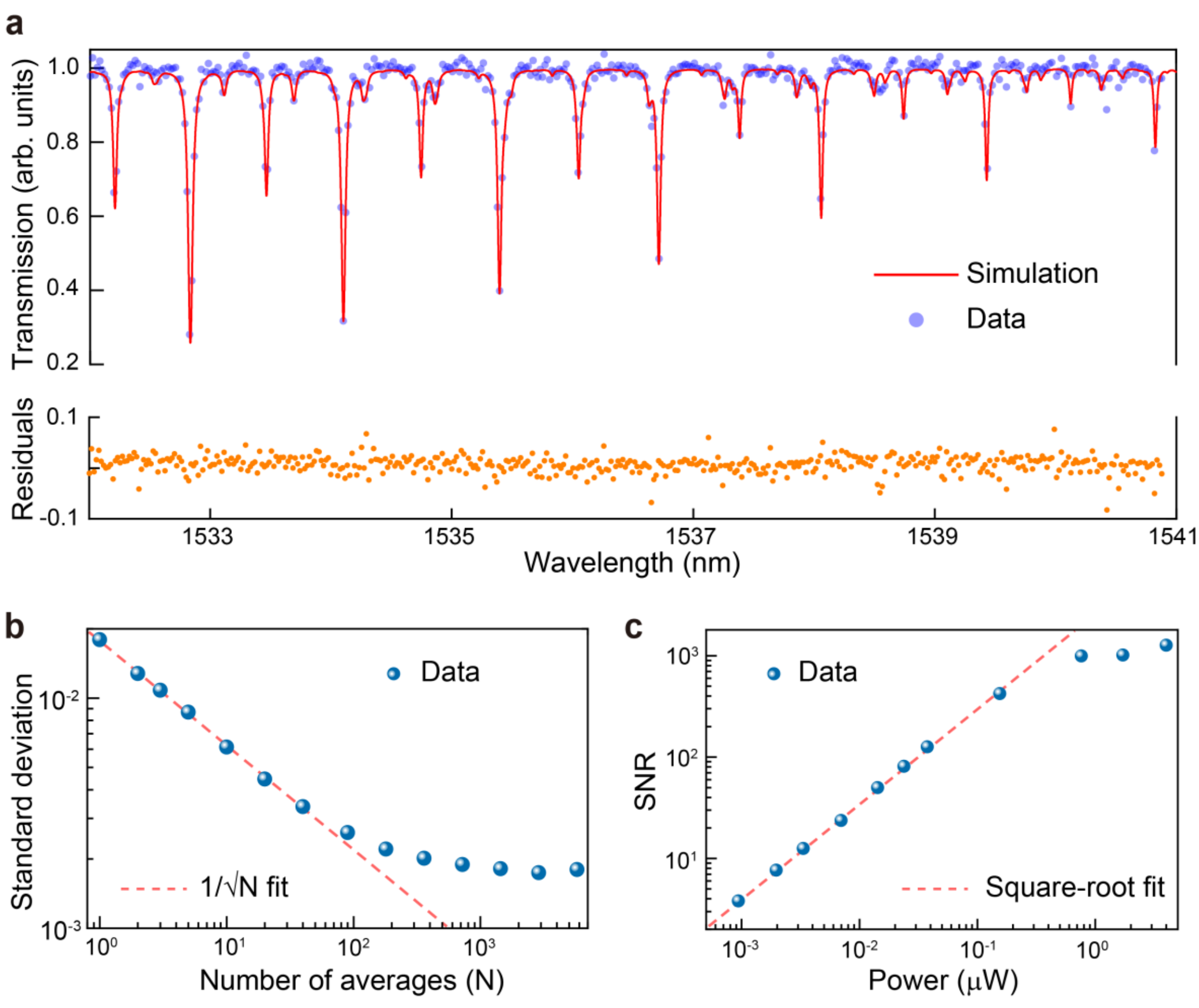

**Fig. 5 | Spectral measurements under scattering conditions.**

**a,** Reconstructed spectrum obtained through scattering media, along with corresponding simulation results.

**b,** Standard deviation of the residuals in **a** as a function of the averaging number, measured at an average detected power of 4 μW.

**c,** Signal-to-noise ratio (SNR) versus optical power for 200 averaging cycles.

Beyond its data-compression benefits, the intensity-only nature of our reconstructive approach provides the unique advantage of inherent immunity to phase scrambling. We demonstrate this robustness in highly scattering environments. As indicated by the black dashed box in Fig. 1a, a 2-mm-thick frosted-glass plate of 80% transmittance is placed in front of the gas cell. By scrambling the phase of the transmitted comb modes, this diffuser renders the coherent heterodyne detection required for methods such as dual-comb spectroscopy ineffective. In contrast, our computational-sensing scheme remains robust under such conditions because it relies only on optical power or intensity. Figure 5a shows the reconstructed spectrum (purple dots) obtained using differential Hadamard encoding, overlaid with the corresponding HITRAN simulation (red curve). The residuals (orange dots) exhibit a SD of 1.7%, which decreases further with data averaging (Fig. 5b), following the same scaling behaviour observed in Fig. 3c. These results demonstrate that our method enables reliable, high-resolution spectroscopic sensing behind strongly scattering media, with broad applicability to environmental monitoring [3, 6].

Moreover, we quantify the signal-to-noise ratio (SNR) as the reciprocal of the SD, which increases with the square root of the received average optical power (P), as shown in Fig. 5c (each data point corresponds to 200-fold averaging). For $P > 1$ μW, the SNR saturates due to the limited dynamic range of our detection and digitization system. Nevertheless, this $\sqrt{P}$ scaling is consistent with shot-noise-limited detection [29].

## Measurement with single-photon sensitivity

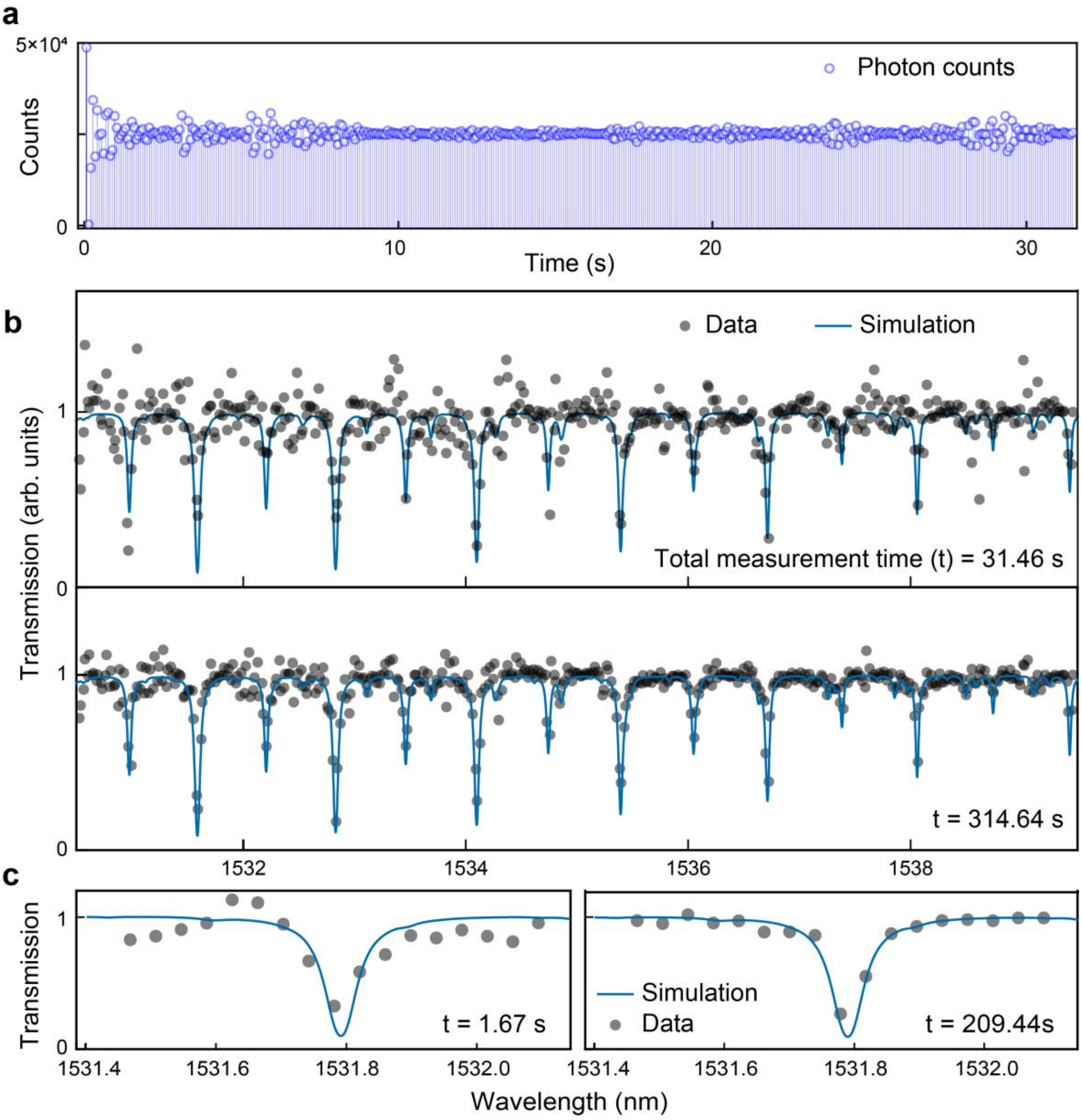

**Fig. 6 | Single-photon spectral measurements.**

**a,** Time-domain histogram from the single-photon detector under gated acquisition.

**b,** Broadband spectral reconstruction at a detected power level of 46 fW.

**c,** Narrowband spectral reconstruction with 5.3 fW of detected power.

Finally, we demonstrate non-cooperative spectroscopic sensing in a photon-starved regime. In this experiment, we extend the sensing distance to beyond 10 m. A collimated comb beam (10 mW total power, corresponding to an average of 0.04 mW per mode) illuminates a distant wall, and the diffusely scattered light is subsequently collected by a telescope (Fig. 2a). The returned photons are detected by a single-photon detector and recorded using a photon counter. Figure 6a shows the time-domain photon-counting histogram measured at a photon flux of $3.6\times10^{5}$ photons $s^{-1}$ (corresponding to 46 fW of detected power or $1.3\times10^{-3}$ photons per pulse). From these data, we reconstruct the mode-resolved absorption spectrum over 1530.5–1539.5 nm with a total acquisition time of 31.46 s (Fig. 6b). For comparison, a higher-SNR spectrum obtained with an extended acquisition time of 314.64 s is also shown. The spectral SNR also scales as $\sqrt{t}$.

We note that restricting the measurement to a narrower spectral window substantially reduces the required photon flux for accurate spectral reconstruction, as spectral filtering effectively isolates comb photons from ambient background light. For instance, we reduce the comb spectral bandwith to 1 nm and measure the $\nu_1+\nu_3$ band P(11) transition of $^{12}C_2H_2$ under a photon flux of only $4.1\times10^{4}$ photons $s^{-1}$, corresponding to 5.3 fW of detected power or $1.5\times10^{-4}$ photons per pulse. The reconstructed spectra for total measurement times of 1.67 s and 209.44 s (Fig. 6c) match the simulated spectra, confirming that reliable spectroscopic retrieval is achievable even in the few-photon regime.

## Discussion

Recently, single-photon DCS has been demonstrated for low-light detection in several works [30-37]. A particularly relevant advance is single-photon dual-comb spectral ghost imaging [36], which retrieves spectra by correlating dual-comb reference signals with single-photon counts. However, these approaches rely on mutually coherent dual combs and phase-sensitive multi-heterodyne detection performed under controlled environments. They also require acquiring dual-comb interferograms over long time windows, yielding substantial data volumes that burden real-time processing [38]. In contrast, our approach operates without reference spectra and employs a single mode-programmable comb with intensity-only detection, enabling data-efficient, broadband, mode-resolved spectroscopy under non-cooperative conditions.

In fact, combining a comb with a VIPA-based 2D dispersive spectrometer has long enabled high-precision, mode-resolved spectroscopy [21]. As in these works, the spectral resolution of our system is set by the comb mode spacing, and ultimately by the comb linewidth, rather than by the dispersive optics, provided that individual comb modes are fully resolved. Also, the absolute mode frequencies can be retrieved with high accuracy. In our setup, the frequency accuracy is currently limited by the wavemeter (±10 MHz) used to calibrate the EO-comb seed laser. In principle, this can be improved to the intrinsic linewidth of the seed laser (<0.1 kHz), corresponding to a fractional uncertainty of $5\times10^{-12}$ at $f_{cw}$=193.4 THz. Distinct from prior VIPA–comb implementations, our scheme integrates the 2D disperser with a DMD for encoding the comb modes, allowing the use of a highly sensitive single-pixel detector in place of a camera. This architecture enables compressed sensing and single-photon sensitivity, both of which are crucial for time-consuming, low-light measurements.

These capabilities provide clear advantages for real-world gas-sensing applications. In particular, by retaining the metrological strengths of frequency comb spectroscopy, the method is promising for precise identification and quantification of multiple gas species. Moreover, its ability to operate without target-side reflectors makes it uniquely suited for standoff sensing in hazardous or unstructured environments, including industrial leak localization, explosive chemical accident sites, and open-air paths where only diffuse returns are available. Furthermore, we foresee this technique being integrated into intelligent sensing platforms, including drones [39] and satellites [40], thereby significantly boosting their spectral analysis capabilities for remote monitoring of greenhouse gases and pollutants. Finally, we note that when combined with sensitivity-enhanced techniques [41-43], such as cavity-enhanced spectroscopy [44-46] or photoacoustic spectroscopy [47-50], our method could also deliver high sensitivity for trace-gas analysis, although this aspect is not the focus of the present work.

In summary, we introduce a computational spectroscopic technique based on a mode-programmable comb, which combines high spectral resolution with broad simultaneous bandwidth (10 nm or 1.27 THz without spectral stitching), single-photon sensitivity, and the compressed, non-cooperative sensing ability. These capabilities position our approach as a powerful tool for widespread gas sensing applications.

## Methods

### EO comb generation

For broadband comb generation, a 1550-nm continuous-wave (CW) laser (Adjustik E15, NKT Photonics; linewidth < 0.1 kHz; output power 40 mW) is injected into a Fabry–Perot electro-optic (FP-EO) comb generator (WTEC-01, OptoComb). The FP-EO modulator is driven by a 5-GHz sinusoidal signal from a signal generator (SMB100A, Rohde & Schwarz) referenced to a rubidium frequency standard. The FP-EO comb output is subsequently amplified to 500 mW using a bidirectionally pumped erbium-doped fiber amplifier (EDFA) and spectrally broadened in a 100-m-long dispersion-managed highly nonlinear fiber (HNLF). The HNLF exhibits negative dispersion at 1550 nm with a dispersion slope of 0.03 ps $nm^{-2}$ $km^{-1}$, an attenuation below 1.5 dB $km^{-1}$, and a nonlinear coefficient exceeding 10 $W^{-1}$ $km^{-1}$. Fine tuning of the comb mode positions is achieved by adjusting the CW laser frequency via a piezoelectric transducer and stabilizing the FP-EO cavity temperature, enabling discrete frequency shifts with a step size of 100 MHz for high-resolution measurements.

### VIPA-based 2D disperser

The two-dimensional (2D) disperser comprises a virtually imaged phased array (VIPA; OP-6721-1686-8, LightMachinery) in combination with a transmission grating (T-966C-16×10-94, LightSmyth). The VIPA provides a spectral resolution of 730 MHz and a free spectral range (FSR) of 60 GHz over the wavelength range of 1.5–1.7 μm. When the input spectral bandwidth exceeds the VIPA FSR, multiple diffraction orders spatially overlap at the VIPA output. To separate these orders, a transmission grating with 966 lines $mm^{-1}$ is oriented orthogonally to the VIPA dispersion axis. The grating is illuminated at an incidence angle of 48°, yielding a spectral resolution of 13.4 GHz, which is finer than the VIPA FSR. As a result, the combined 2D disperser fully resolves 5-GHz-spaced comb modes in two-dimensional space.

### Spectrum encoding and reconstruction

In this proof-of-concept demonstration, a 256 × 256 Hadamard matrix is employed to encode 228 comb teeth. The first 228 columns of the matrix correspond to comb-line-resolved spatial spots, while the remaining 28 columns are assigned to unused (dark) regions of the digital micromirror device (DMD). Spectral reconstruction is performed using different strategies depending on the number of measurements $m$. When the reconstruction dimension satisfies $n = m = N$, where $N$ is the Hadamard encoding dimension, the spectrum is recovered via the inverse Hadamard transform. For the overdetermined case ($n \leq m < N$), the spectrum is reconstructed by solving a least-squares problem using the singular value decomposition (SVD)-based Moore–Penrose pseudoinverse. In the underdetermined regime ($m < n$), total-variation (TV) regularization is employed, and the reconstruction is carried out using the TVAL3 algorithm. The normalized transmission spectrum is obtained by dividing the reconstructed gas-absorption signal by the reconstructed background signal.

**Data acquisition**

In the experiment, the encoded comb is split into a reference beam and a measurement beam using a 50:50 fiber coupler. The measurement beam propagates through the target gas, whereas the reference beam is directly detected to provide a background signal. As shown in Fig. 2a, both beams are measured using single-pixel detectors. The detector outputs are acquired simultaneously with a 12-bit digital oscilloscope (SDS1202X, Siglent) at a sampling rate of 100 kSa/s.

## Acknowledgements

This work is financially supported by Quantum Science and Technology-National Science and Technology Major Project (2023ZD0301000) and Scientific Research Innovation Capability Support Project for Young Faculty with grant no. ZYGXQNJSKYCXNLZCXM-E8.

## Author contributions

W. C. initiated and conceived the idea. M. Y. and H. Z. contributed to the development of the concept. M. Y., Y. L., and Z. W. designed the experiments. D. Z., X. M. and Z. W. conducted the experiment. Y. C., Mei Y., and M. L., built the comb source. Z. J. W. and X. Z. optimized the setup. D. Z., Z. W., W. C., Y. Y., and C. S. developed the code and analyzed the data. M. Y. and Z. W. drafted the manuscript. W. C., H. L., K. H., and H. Z. revised the manuscript. All authors provided comments and suggestions for improvements.

## Competing interests

The authors declare no competing interests.

## Data availability

The data that support the findings of this study are available from the corresponding author upon reasonable request.

## Author information

Correspondence and requests for materials should be addressed to M. Y. (myan@lps.ecnu.edu.cn), Y. L. (yanliang@usst.edu.cn), or W.C. (cweiwei@sjtu.edu.cn).